\documentclass[11pt,a4paper]{article}
\usepackage{amsfonts}%
\usepackage{amsmath}%
\usepackage{amssymb}%
\usepackage{graphicx}
\newcommand{\bell}{\mathbf{\ell}}

\newcommand{\D}{\mathfrak{D}}

\newcommand{\half}{{\textstyle \frac{1}{2}}}

\newcommand{\bx}{\mathbf{x}}
\newcommand{\by}{\mathbf{y}}
\newcommand{\bb}{\mathbf{B}}

\newcommand{\bzero}{\mathbf{0}}
\newcommand{\Order}{\mathcal{O}}

\newcommand{\bel}[1]{\begin{equation}\label{#1}}
\newcommand{\bal}[1]{\begin{eqnarray}\label{#1}}
\newcommand{\ee}{\end{equation}}
\newcommand{\ea}{\end{eqnarray}}
\newcommand{\equ}[1]{~Eq.(\ref{#1})}

\renewcommand{\P}{\mathcal{P}}
\renewcommand{\H}{\mathfrak{H}}
\newcommand{\C}{\mathfrak{C}}

\newcommand{\vev}[1]{\langle #1\rangle}
\newcommand{\drop}[1]{}

\begin{document}

\begin{center}
{\Large Local Measures of Convex Surfaces induced by the Wiener Measure of Paths}\\~\\
Martin Schaden\\
Rutgers University, 101 Warren Street, Newark NJ 07102
\end{center}
\begin{abstract}
The Wiener measure induces a measure of closed, convex, $d-1$-dimensional, Euclidean (hyper-)surfaces that are the convex hulls of closed $d$-dimensional Brownian bridges. I present arguments and numerical evidence that this measure, for odd $d$, is generated by a local classical action of length dimension~2 that depends on geometric invariants of the $d-1$-dimensional surface only.
\end{abstract}



\section{Introduction}\label{aba:sec1}
It was recently\cite{Schaden09a} observed that the spectral function $\phi_\D(\beta)$, or trace of the heat kernel $\mathfrak{K}_\D(\beta)$, of a free massless scalar field that vanishes on the boundary $\partial \D$ of a $d$-dimensional convex domain $\D$ ,
\bel{spfunc}
\phi_\D(\beta)=\sum_{n\in\mathbb{N}} e^{-\beta\lambda_n/2}=\int_\D d\bx \mathfrak{K}_\D(\bx,\bx;\beta),
\ee
naturally induces a measure on convex surfaces that are the convex hulls of closed Brownian bridges. Here $\{\lambda_n, n\in\mathbb{N}\}$ is the spectrum of the (negative) Laplacian $\Delta$ with Dirichlet boundary conditions on  $\partial\D$ and the heat kernel $\mathfrak{K}_\D(\bx,\by;\beta)$  solves the diffusion equation, $\partial_\beta\mathfrak{K}_\D(\bx,\by;\beta)=\half\Delta_x \mathfrak{K}_\D(\bx,\by;\beta)$,
with initial condition $\mathfrak{K}_\D(\bx,\by;0)=\delta(\bx-\by)$ and vanishes on $\partial\D$.

The Feynman-Kac theorem\cite{F48Kac66} implies that $\phi_\D(\beta)$ can be expressed by the probability $\P[\bell_\beta(\bx)\subset\D ]$ that a standard Brownian bridge (SBB) $\bell_\beta(\bx)=\{\bx+\bb_\tau, 0\leq \tau\leq \beta;
\bb_0=\bb_\beta=\bzero\}$ starting at $\bx$ and returning to $\bx$ in
"proper time" $\beta$ does not exit $\D$,
\bel{support}
\phi_\D(\beta)=\int_\D \frac{d\bx}{(2\pi \beta)^{d/2}} \P[\bell_\beta(\bx)\subset\D ]\ .
\ee
A numerical approximation to a SBB derives from a discrete random walk $\{\mathbf{X}_n, n=0,\dots, N; \mathbf{X}_0=\mathbf{0}\}$ of $N$ independent steps with average displacement $\vev{\mathbf{X}_{n+1}-\mathbf{X}_{n}}=\mathbf{0}$ and variance $\vev{(\mathbf{X}_{n+1}-\mathbf{X}_{n})^2}=\beta d /N$. $\bell_\beta(\bx)$ then is approximated by $\{\mathbf{Y}_n, n=0,\dots, N; \mathbf{Y}_n=\mathbf{X}_n+\bx-n \mathbf{X}_N/N\}$. $\mathbf{Y}_N=\mathbf{Y}_0=\bx$ and the steps remain uncorrelated and of variance $d\beta/N$ by construction. The continuum limit is obtained by letting  $N\rightarrow\infty$.

To ascertain whether a SBB is entirely within a bounded domain generally requires examining all of its $N$ points.  However, in the case of \emph{convex} bounded domains $\D$, it suffices to check whether the smallest convex surface enclosing all points of the SBB, its \emph{convex hull}, lies entirely within $\D$.  Of advantage is that the convex hull of a SBB has exponentially fewer vertices (see Fig.1) than the original loop has points and that its construction is independent of the convex domain $\D$. One thus can generate and store a large number of convex hulls of SBBs once and for all and only has to check whether the few vertices of the hulls lie within a particular convex domain $\D$.  The computer time required to construct the hull of a set of $N$ points is $\Order(N\ln N)$ only\cite{Mount02} and thus is comparable to the time required to examine the $N$ points of a loop -- but the convex hull can be used for any convex domain. The ensemble of convex hulls of SBBs furthermore shares all global symmetries of the Wiener measure.
\newcounter{figures}
\begin{figure}
\refstepcounter{figures}\label{fig1}
\includegraphics[width=2.5in]{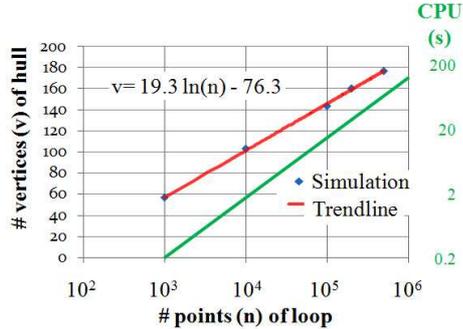}
\caption{Average number of vertices $v$ (left axis) of the hull of a 3-dimensional SBB with $n$ points (red). The trendline is given by the inset formula.  The right axis and lower (green)
line give the average CPU-time required to compute the hulls on a laptop with a 2GHz processor.}
\label{aba:fig1}
\end{figure}
\section{Measuring convex surfaces with the Wiener measure}
One can define open sets of convex domains $\C=\{C \textrm{ is convex}\}$ that include all convex domains that, much like russian dolls, fit into a particular one. One can similarly define open sets of the boundaries of convex domains,
\bel{openset}
{\cal M}(\D):=\{ \partial C; C\subset\D\in\C\}=\{ {\cal H}[\bell_\beta(\bx)], \bell_\beta(\bx)\subset \D \}\ ,
\ee
where ${\cal H}[\bell_\beta(\bx)]$ is the convex hull of the SBB $\bell_\beta(\bx)$ based at $\bx\in\D$. In view of the correspondence in \equ{openset} it is natural to assign the set ${\cal M}(\D)$ the translation, $O(d)$ and scale- invariant measure provided by the spectral function,
\bel{measure}
\mu_\beta [{\cal M}(\D)]:=\phi_\D(\beta)=\int_\D \frac{d\bx}{(2\pi\beta)^{d/2}} \P[\bell_\beta(\bx)\subset\D]\ .
\ee
Although $\tilde{\cal M}(\D)$ does not refer to a scale, the induced probability measures on convex surfaces defined by \equ{measure} depend on the parameter $\beta$, of length dimension $2$, that characterizes the SBBs. The measures $\mu_\beta$ may be extended to the entire Borel set by using the definition of \equ{measure} and properties of the probability space of Brownian loops\cite{Oksendahl87} as well as the fact that $\bigcap_{i>0}{\cal M}(\D_i)={\cal M}(\cap_{i>0}\D_i)$ .

\section{Locality}
These  measures on convex surfaces are \emph{local} in the sense that they can be generated by a Metropolis-like algorithm\cite{MHalgorithm} involving only local updates of convex surfaces. This perhaps is less surprising upon recalling that convexity is equivalent to demanding that the scalar curvature of the surface is positive everywhere.  The latter constraint is \emph{local}. The following construction of the ensemble of convex surfaces with a measure proportional to \equ{measure} is \emph{not} the most efficient, but does imply that these measures on convex surfaces in $d>1$ are generated by a local action.

Note first that the Wiener measure is local in this sense, because the relative frequency of two SBBs, $\bell'_\beta$ and $\bell_\beta$, that differ only in the immediate vicinity of  $\mathbf{Q}=\bx_k$ is,
\bal{Wiener}
\frac{d\P[\bell'_\beta]}{d\P[\bell_\beta]}&\sim & 1+N (\bx'_k-\bx_k)\cdot(\bx_{k+1}-2\bx_k+\bx_{k-1})/\beta\nonumber\\
& = & 1+\delta_Q S\textrm{   with   } S=\int_0^\beta  \hspace{-.5 em}{\dot\bx}^2 d\tau;\ \ \bx(\beta)=\bx(0)
\ea
The local classical action $S$ in this case may be reconstructed explicitly, and a Metropolis algorithm\cite{MHalgorithm} may be designed that proposes local updates of the SBB with precisely this probability, i.e. the algorithm accepts every proposal.

Now consider the corresponding two hulls $h={\cal H}[\bell_\beta]$ and $h'={\cal H}[\bell'_\beta]$. There are four distinct possibilities:(i) $\bx_k$ is a vertex of $h$ and $\bx'_k$ is a vertex of $h'$; (ii) $\bx_k$ is a vertex of $h$ but $\bx'_k$ is not a vertex of $h'$; (iii) $\bx_k$ is not a vertex of $h$ but $\bx'_k$ is a vertex of $h'$ and (iv) $\bx_k$ is not a vertex of $h$ and $\bx'_k$ is not a vertex of $h'$.
Since only a single point of the underlying loop is moved, (iv) implies that $h'=h$. In all other cases, $h'\neq h$ and at least one vertex has changed. The update is almost local in that the average distance $\delta$ from the point $\mathbf{Q}=(\bx_k+\bx'_k)/2$ on the convex surface that is affected is of order $\delta\sim\sqrt{2\bar R \Delta}=\Order(\sqrt{\beta} N^{-1/4})$. [The average radius of curvature  of the convex surface $\bar R=\Order(\sqrt{\beta})$ does not depend on $N$ and the average displacement of the updated point is $\Delta\bx=\sqrt{\vev{(\bx'-\bx)^2}}=\Order(\sqrt[\beta/N])$.] This region thus decreases with increasing refinement $N$ of the SBB like $N^{(1-d)/4}$ and the update becomes local in the continuum limit. As can be seen in Fig.~1 the number of vertices $v$ of the hull of a SBB is proportional to $\ln(N)$ and thus, for large $N$, almost always at most one vertex of the hull is updated. For large $N$ case~(iv) occurs most frequently and an algorithm that updates large sections of the underlying loop is far more efficient than this local one.

However, I argue that the possibility of constructing a local algorithm implies that rare events are generated by a local classical action. Since the probability for generating a SBB of large extent $s$ decreases exponentially $\sim \exp[\Order(s^2/\beta)]$, the probability for generating a convex hull much larger than average should also asymptotically decrease as
\bel{asymptotic}
\P[\D\subset\H[\bell_\beta]]\vbox{\hbox{\vspace{-2ex}$\genfrac{}{}{0pt}{}{\longrightarrow}{\beta\sim 0}$}}\propto e^{-S_d(\D)/\beta}\ ,
\ee
where $S_d(\D)$ is a classical action of length dimension $2$ that can only depend on geometrical characteristics of $\D$. Locality of the measure  in the continuum limit implies that $S_d(\D)$ is an integrated density that depends on the metric and its derivatives only. The local constraint that $\D$  have positive curvature everywhere is not holomorphic and therefore is not reflected in the form of $S_d(\D)$, much as the lower bound for the motion of a bouncing ball does not appear in its action.
\section{Numerical results}
$\mathbf{d=1}$ The convex hull of a SBB in 1~dimension consists of two points, the maximum and minimum values of the SBB. The only translation invariant action of length dimension two is the square of the distance between these. Explicit calculation gives\cite{Schaden09b},
\bel{1dim}
\P[|\H[\bell_\beta]|<s]=1-2\sum_{n=1}^\infty ((2 s n)^2/\beta-1)e^{-(2 s n)^2/(2\beta)}\vbox{\hbox{\vspace{-2ex}$\genfrac{}{}{0pt}{}{\longrightarrow}{\beta\sim 0}$}}1-8 s^2/\beta e^{-2 s^2/\beta},
\ee
where $|\H[\bell_\beta]|$ denotes the extent of the convex hull, that is the distance between its two points. The asymptotic behavior of \equ{1dim} has the expected form with an action $S_1([a,b])=2 (a-b)^2$ that depends on the two points of the hull only. However, $S_1(\D)$ is not local and the correlation matrix is constant due to the failure of the previous geometrical argument for a discrete set of points. As shown in the talk, the agreement of numerical simulations with \equ{1dim} is excellent.
\newline $\mathbf{d=2}$ The convex hull of a SBB in two dimensions numerically is a piecewise linear, closed, one-dimensional curve, whose length may be called the perimeter of the SBB. It is a local geometrical invariant of dimension~1. \equ{asymptotic} suggests that the measure for rare events will not scale with the perimeter (see Fig.~2), but rather with the area enclosed by it, $ S_2(\D)\propto A=\int_\D d^2 x $ . The measure also scales with the local quantity of dimension~2 that is the sum of the squares of the lengths of the piecewise linear sections of the curve $P^2=\sum_i s_i^2$, but  $P^2\rightarrow 0$ as $N\rightarrow \infty$ and does not survive the continuum limit. The numerical evidence shown in Fig~2a supports this identification.
\begin{figure}
\refstepcounter{figures}\label{fig2}
\includegraphics[width=4.5in]{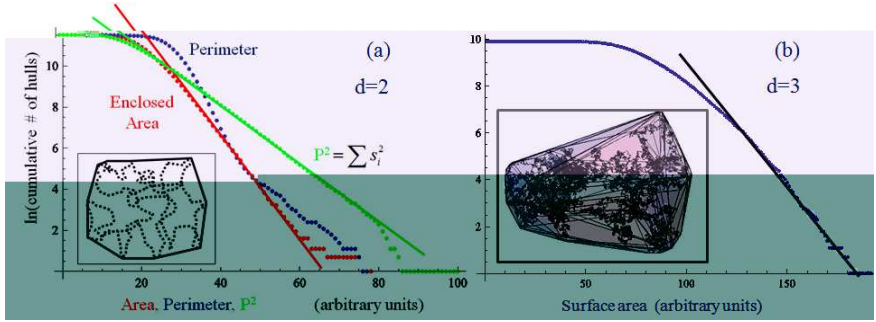}
\caption{Logarithm of the cumulative statistics (a) for the enclosed area (red), the perimeter (blue) and the invariant $P^2$ (green) of the convex hulls of $10^5$ SBBs with $N=10^4$ points in $d=2$ dimensions and (b) the area of the convex hulls of $10^4$ SBBs with $N=10^4$ steps in $d=3$ dimensions. The exponential decrease in rare events with the enclosed area and with $P^2$ in $d=2$ and with the surface area in $d=3$ is indicated. The insets show sketches of SBBs and their convex hulls in $d=2$ and $d=3$ dimensions.}
\label{aba:fig2}
\end{figure}
\newline $\mathbf{d=3}$ The convex hull of a SBB in $d=3$ is a triangulated 2-dimensional surface. The only local action of length dimension~2 is its area $S_3(\D)\propto \int_{\partial D} d^2x \sqrt{\det{g_{ik}(\bx)}}$, where  $g_{ik}(\bx)$ is the metric \emph{on} $\partial\D$ (see Fig.~2b).
\section{Outlook and a conjecture for $d>3$ dimensions}
Numerical evidence in four and higher dimensions were not presented, because the construction of hulls and local actions in this case is considerably more involved and has no direct application to Casimir effects. A dimensional argument, however, suggests that the local classical action for \emph{even}-dimensional convex surfaces is determined by the metric on the surface and its (tangential) derivatives only.  Local classical actions that asymptotically describe the measures for \emph{odd}-dimensional convex surfaces on the other hand necessarily depend on the bulk metric (see Fig.~2a), since the number of derivatives in local geometric invariants is always even. One thus is led to conjecture that  $S_5(\D)\propto \int_{\partial D} d^4x \sqrt{\det{g_{ik}(\bx)}} R(\bx)$, may be intrinsically defined in terms of the metric and associated Ricci curvature scalar $R(\bx)$ of the 4-dimensional convex hyper-surface embedded in 5-dimensional flat Euclidean space. The nontrivial intrinsic local classical action in odd $d=7,9,\dots$ are sums of higher derivative terms that are not uniquely determined by dimensional considerations alone.
\section*{Acknowledgments} I would like to thank the organizers of QFEXT09 for superbly managing a very engaging conference. This work was supported by the National Science Foundation with Grant No. 0902054.

\end{document}